\documentclass[prb,twocolumn,aps,10pt]{revtex4-1}

\usepackage{epsfig}
\usepackage{color}
\usepackage{amsmath}

\renewcommand{\vec}[1]{{\rm\bf #1}}
\renewcommand{\Re}{\mathop{\mathrm{Re}}\nolimits}
\renewcommand{\Im}{\mathop{\mathrm{Im}}\nolimits}

\newcommand{\ep}{\epsilon}
\newcommand{\sign}{\mathop{\mathrm{sign}}}

\begin{document}

\title{Effect of anisotropic band curvature on carrier
multiplication in graphene}
\author{D.~M.~Basko}\email{denis.basko@grenoble.cnrs.fr}
\affiliation{Universit\'e Grenoble 1/CNRS, LPMMC UMR 5493,
25 rue des Martyrs, 38042 Grenoble, France}

\begin{abstract}
We study relaxation of an excited electron in the conduction
band of intrinsic graphene at zero temperature due to
production of interband electron-hole pairs by Coulomb
interaction.
The electronic band curvature, being anisotropic because of
trigonal warping, is shown to suppress relaxation for a range
of directions of the initial electron momentum. For other
directions, relaxation is allowed only if the curvature
exceeds a finite critical value; otherwise, a nondecaying
quasiparticle state is found to exist.
\end{abstract}

\maketitle

\section{Introducton}

Carrier multiplication is a process in which a single photon,
absorbed by a material, produces several electron-hole (e-h)
pairs.
Typically, this happens when the primary photoexcited e-h
pair produces a number of secondary pairs of smaller energy
via electron-electron collisions.
This process is very important for optoelectronic
applications: the more e-h pairs are produced by a single
photon, the more efficiently one can convert light into
electric current.
Graphene is an obvious candidate for efficient carrier
multiplication, since (i)~it has wide electronic bands and
no energy gap, and (ii)~electron-electron scattering can be
much faster than electron-phonon scattering.
Indeed, the dynamics of photoexcited carriers in graphene
has become a subject of many studies, both theoretical
and experimental.%
\cite{Rana2007,George2008,Dawlaty2008,Sun2008,Bao2009,Kumar2009,%
Newson2009,Plochocka2009,Lui2010,Winzer2010,Hale2011,%
Winnerl2011,Breusing2011,Kim2011,Strait2011,Li2012,Winzer2012,%
Song2012,Brida2012,Tielrooj2012,Winzer2013,Winnerl2013,SunWu2013}

\begin{figure}
\includegraphics[width=8cm]{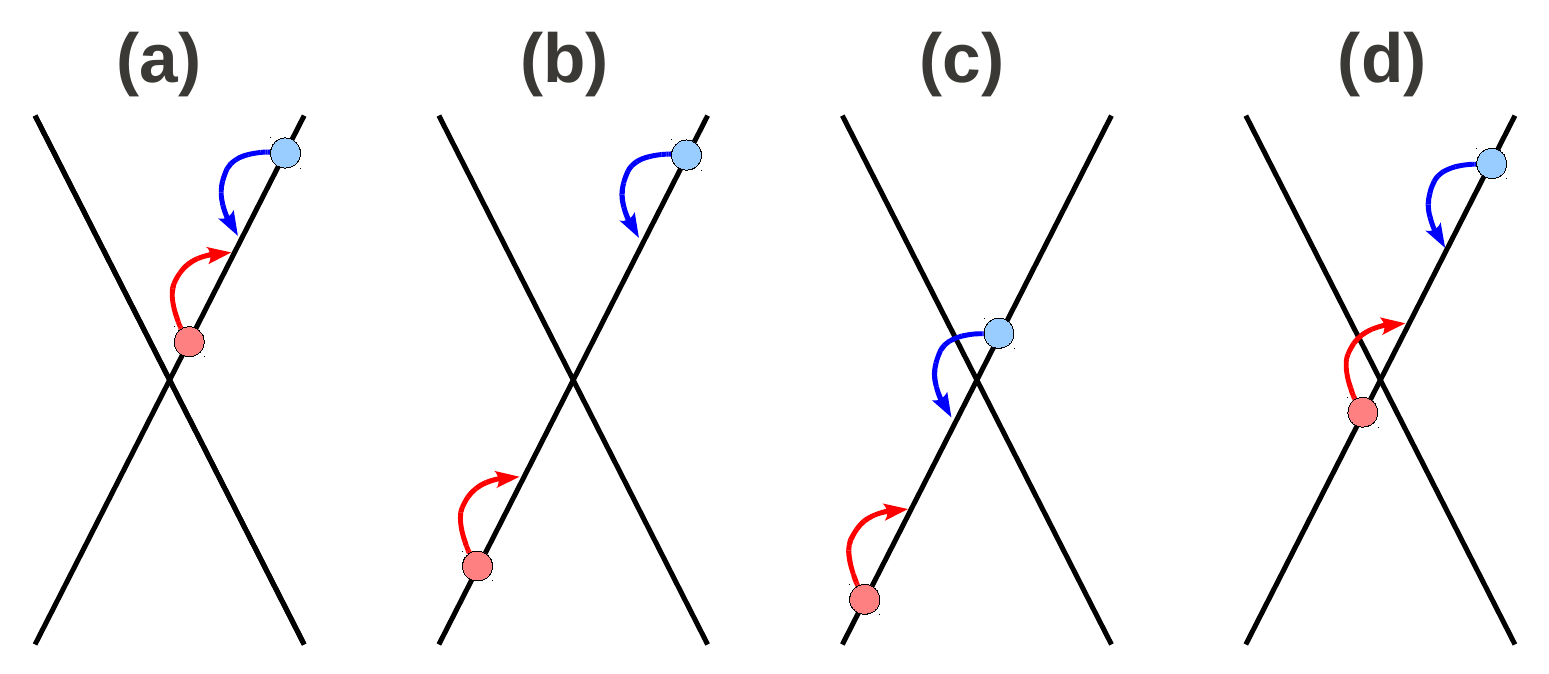}
\caption{\label{fig:collisions}
Various collision processes for an excited electron
in the conduction band. The straight lines
represent the Dirac cones and the circles represent electrons.
Processes (a),~(b) conserve the number of carriers in the
conduction and valence bands separately. Processes (a)--(c)
are forbidden by the Pauli principle in the intrinsic graphene
at zero temperature.
Process~(d) corresponds to carrier multiplication (the initial
electron produces one more electron and one hole).
}
\end{figure}

In spite of the numerous studies, this dynamics is still not
fully understood. Notably, the most basic issue, that of the
role of electron-electron collisions in the relaxation of a
single photoexcited carrier in the intrinsic graphene, is still
under debate. It is well known (see, e.~g., Ref.~\onlinecite{LL9})
that due to simultaneous energy and momentum conservation, decay
of quasiparticles is allowed or forbidden,
depending on the curvature of the quasiparticle spectrum.
In the context of graphene, the Dirac spectrum is linear, so
it is exactly on the borderline between the two
cases.\cite{Gonzalez1996} Indeed, if an electron in the conduction
band with momentum~$\vec{p}$ and energy $v|\vec{p}|$ ($v$~being
the Dirac velocity) is scattered into the state with another
momentum~$\vec{p}'$ and lower energy $v|\vec{p}'|<v|\vec{p}|$,
creating a hole in the valence band with momentum~$\vec{p}_h$
and another electron in the conduction band with momentum
$\vec{p}_e$, the momentum and energy conservation conditions,
\begin{subequations}\begin{eqnarray}
&&\vec{p}=\vec{p}'+\vec{p}_e+\vec{p}_h,\\
&&v|\vec{p}|=v|\vec{p}'|+v|\vec{p}_e|+v|\vec{p}_h|,
\end{eqnarray}\end{subequations}
are compatible only in the special case when all vectors lie on
the same line. Different ways to resolve the
uncertainty have been advocated.\cite{Rana2007,Li2012,Brida2012,%
Winzer2013,Fritz2008,Foster2009,Basko2009,Golub2011}
Note that the collision process in question is precisely the
one responsible for carrier multiplication
(Fig.~\ref{fig:collisions}).

It might seem that importance of the above-mentioned problem of
a single excited carrier is limited to low photoexcitation
intensities. However, if many e-h pairs are created under intense
photoexcitation, it is important to know whether the
population equilibration between the valence and the conduction
band happens as quickly as the thermalization within each band.
For example, only if the interband population relaxation is slow
enough, a population inversion between the 
bands can be achieved, and one can think about lasing.
Among various two-electron collision processes, shown in
Fig.~\ref{fig:collisions}, only the processes (d)~(carrier 
multiplication) and (c)~(Auger recombination, reciprocal
to the multiplication) can transfer electrons between the bands.
If these are suppressed, three-particle collisions are
required to equilibrate the populations of the conduction and
the valence bands.

In the present work, we study relaxation of an excited electron
in the conduction band of intrinsic graphene at zero
temperature due to electron-electron collisions (the process~(d)
in Fig.~\ref{fig:collisions}), going beyond the Dirac approximation
and taking into account the electronic band curvature. Because the
curvature is anisotropic due to the trigonal warping, the
result turns out to depend on the direction of the initial electron
momentum, as was also noted in Ref.~\onlinecite{Golub2011}.
For a certain range of directions, the process is forbidden. For
the directions when relaxation is allowed, we calculate its rate.

If the curvature is weak, the problem corresponds to that of a
discrete state coupled to a continuum whose density of states
is abruptly cut off precisely at the energy of the discrete
level, which can be viewed as a special case of the Fano
problem.\cite{Fano}
Indeed, in the sector with the fixed total momentum~$\vec{p}$,
the single-particle excitation is a discrete level with the
energy $vp$. The three-particle density of states vanishes
below this energy, and exhibits a steplike discontinuity at
the energy $vp$. In this situation, the presumable decay of the
discrete state into the continuum cannot be described by the
Fermi golden rule, since the latter is valid only when the
density of states in the continuum is a smooth function of
energy at the position of the discrete level.
Below it is shown that the quantum-mechanical level repulsion
between the discrete state and the continuum plays a major role
in this problem. As a result, a nondecaying quasiparticle state
exists (that is, its lifetime being determined by mechanisms
other that electron-electron interaction). Still, the
quasiparticle spectral weight is reduced, some part of it being
transferred to the continuum of the multiparticle excitations.
The quasiparticle state can relax by producing electron-hole
pairs only when the band curvature along the allowed directions
exceeds a finite critical value, needed to overcome the level
repulsion.

\section{Calculation}

To derive the results, outlined above, let us describe the
electrons by a two-component column fermionic field operator
$\hat\psi_\alpha(\vec{r})$, where $\alpha=1,\ldots,N$ labels
electronic species (the valley and spin degeneracy in graphene
correspond to $N=4$). The Hamiltonian is given by
\begin{equation}\begin{split}\label{hatH=}
\hat{H}={}&{}
\sum_{\alpha=1}^N\int\hat\psi_\alpha^\dagger(\vec{r})
\,h(-i\boldsymbol{\nabla})\,\hat\psi_\alpha(\vec{r})\,d^2\vec{r}+{}\\
{}&{}+\frac{e^2}{2}\int
\frac{\hat\rho(\vec{r})\,\hat\rho(\vec{r}')}{|\vec{r}-\vec{r}'|}\,
d^2\vec{r}\,d^2\vec{r}'.
\end{split}\end{equation}
The first term in Eq.~(\ref{hatH=}) represents the kinetic energy
of electrons, described by the single-particle Hamiltonian,
\begin{equation}\label{h=}
h(\vec{p})=v(p_x\sigma_x+p_y\sigma_y)
-\zeta_3[(p_x^2-p_y^2)\sigma_x-2p_xp_y\sigma_y]-\zeta_0p^2,
\end{equation}
$\sigma_x,\sigma_y$ being the Pauli matrices. It determines the
single-particle dispersion relation to $O(p^2)$, which we write
as
\begin{equation}\label{epeh=}
\ep_\vec{p}^{e,h}=vp\mp\zeta_0p^2\mp\zeta_3p^2\cos{3}\varphi
\equiv{v}p\pm\zeta_\varphi{p}^2,
\end{equation}
where $\ep^e_\vec{p}(\ep^h_\vec{p})$ is the energy of an
electron in the conduction band (a hole in the valence band)
with momentum~$\vec{p}$, and $\varphi$ is the polar angle
of~$\vec{p}$.
The coefficients can be estimated from the tight-binding
model\cite{Gruneis2008}:
the Dirac velocity $v\approx{7}\:\mbox{eV}\cdot\mbox{\AA}$,
the trigonal warping coefficient
$\zeta_3\approx{5}\:\mbox{eV}\cdot\mbox{\AA}^2$,
and the electron-hole asymmetry
$\zeta_0\approx{0}.8\:\mbox{eV}\cdot\mbox{\AA}^2$.
Neglecting the terms proportional to $\zeta_0,\zeta_3$ in
Eq.~(\ref{h=}) corresponds to the Dirac approximation
$h(\vec{p})=v\vec{p}\cdot\boldsymbol{\sigma}$.

The last term in Eq.~(\ref{hatH=}) describes Coulomb
interaction between the electrons, with the electronic density
\begin{equation}
\hat\rho(\vec{r})=\sum_{\alpha=1}^N
\hat\psi_\alpha^\dagger(\vec{r})\hat\psi_\alpha(\vec{r}),
\end{equation}
and the background dielectric constant $\varepsilon_b$ of
the substrate incorporated into $e^2$. The dimensionless
Coulomb coupling strength $e^2/v$ can be small if
$\varepsilon_b$ is large enough. The largest value of
$e^2/v\approx{2}.2$ is attained for a graphene sheet
suspended in vacuum, $\varepsilon_b=1$ .

The quasiparticle decay rate is given by $-2\Sigma_\vec{p}(\ep)$,
where $\Sigma_\vec{p}(\ep)$, the retarded self-energy projected
on the eigenstate of the single-particle Hamiltonian with
momentum~$\vec{p}$, should be taken at
$\ep$~corresponding to the quasiparticle pole of the Green's
function. In the first approximation, it can be taken on
the mass shell, $\ep=\ep^e_\vec{p}$ or $\ep=-\ep^h_{-\vec{p}}$.
The lowest order of the perturbation theory in the Coulomb
interaction, which contributes to $\Im\Sigma_\vec{p}(\ep)$,
is the second one. It describes decay of the one-particle
state (an electron or a hole) into three-particle excitations
(an electron or a hole plus an e-h pair).
It turns out, however, that in the Dirac approximation, the
second-order $\Im\Sigma_\vec{p}(\ep)$ has a steplike
discontinuity on the mass shell, invalidating the simple
Fermi golden rule
recipe for calculating the decay rate.\cite{Gonzalez1996}
Explicitly, at $|\ep-vp|\ll{vp}$,
\begin{subequations}\begin{eqnarray}\label{ImSigma2=}
&&\Im\Sigma_\vec{p}(\ep)=-\gamma_\vec{p}\,\theta(\ep-vp),\\
&&\gamma_\vec{p}=\pi\left(\frac{N}{24}-\ln\frac{2}{e^{2/3}}\right)
\left(\frac{e^2}v\right)^2vp,\label{gamma=}
\end{eqnarray}\end{subequations}
$\theta(x)$ being the step function. In Eq.~(\ref{gamma=}), 
$N/24$ comes from the bubble diagram (the direct term), while
$\ln(2/e^{2/3})$ comes from the exchange diagram
(see Appendix~\ref{app:ImSigma2} for details).
At $N=4$, the exchange term is more than six times smaller than
the direct term.

One may consider higher orders of the perturbation theory in
$e^2/v$, while remaining within the Dirac approximation.
Decay into $(2n+1)$-particle excitations, which involves
three-particle excitations as virtual intermediate states,
is suppressed by energy and momentum conservation as
$\propto(\ep-vp)^n\theta(\ep-vp)$
(see Appendix~\ref{app:ImSigma2}), so inclusion of many-particle
excitations does not help to resolve the uncertainty.
Dressing the decay into three-particle states by higher-order
corrections can be performed by treating $1/N$ as a formal small
parameter. This selects the random-phase-approximation
(RPA) sequence as the dominant  subclass of diagrams. In RPA,
$\Im\Sigma_\vec{p}(\ep=vp)$ strictly vanishes%
\cite{Gonzalez1996,Khveshchenko2006,DasSarma2007}.
Explicitly (see Appendix \ref{app:ImSigmaRPA}),
\begin{equation}\label{ImSigmaRPA=}
\Im\Sigma_\vec{p}(\ep)=-\frac{4(\ep-vp)}{\pi{N}}\,\theta(\ep-vp)
\ln\frac{\xi_\mathrm{RPA}}{\ep-vp}, 
\end{equation}
where the upper cutoff of the logarithm is
$\xi_\mathrm{RPA}=\min\{vp,(Ne^2/v)^2vp\}$, and
Eq.~(\ref{ImSigmaRPA=}) is valid at $\ep-vp\ll\xi_\mathrm{RPA}$.
If $Ne^2/v\ll{1}$, the perturbative expression of
Eqs.~(\ref{ImSigma2=}),~(\ref{gamma=}) is still valid in the
parametric region of energies $(Ne^2/v)^2vp\ll\ep-vp\ll{vp}$,
while for $Ne^2/v\gtrsim{1}$
Eqs.~(\ref{ImSigma2=}),~(\ref{gamma=}) are never valid, and only 
Eq.~(\ref{ImSigmaRPA=}) holds.

\begin{figure}
\begin{center}
\vspace{0cm}
\includegraphics[width=8cm]{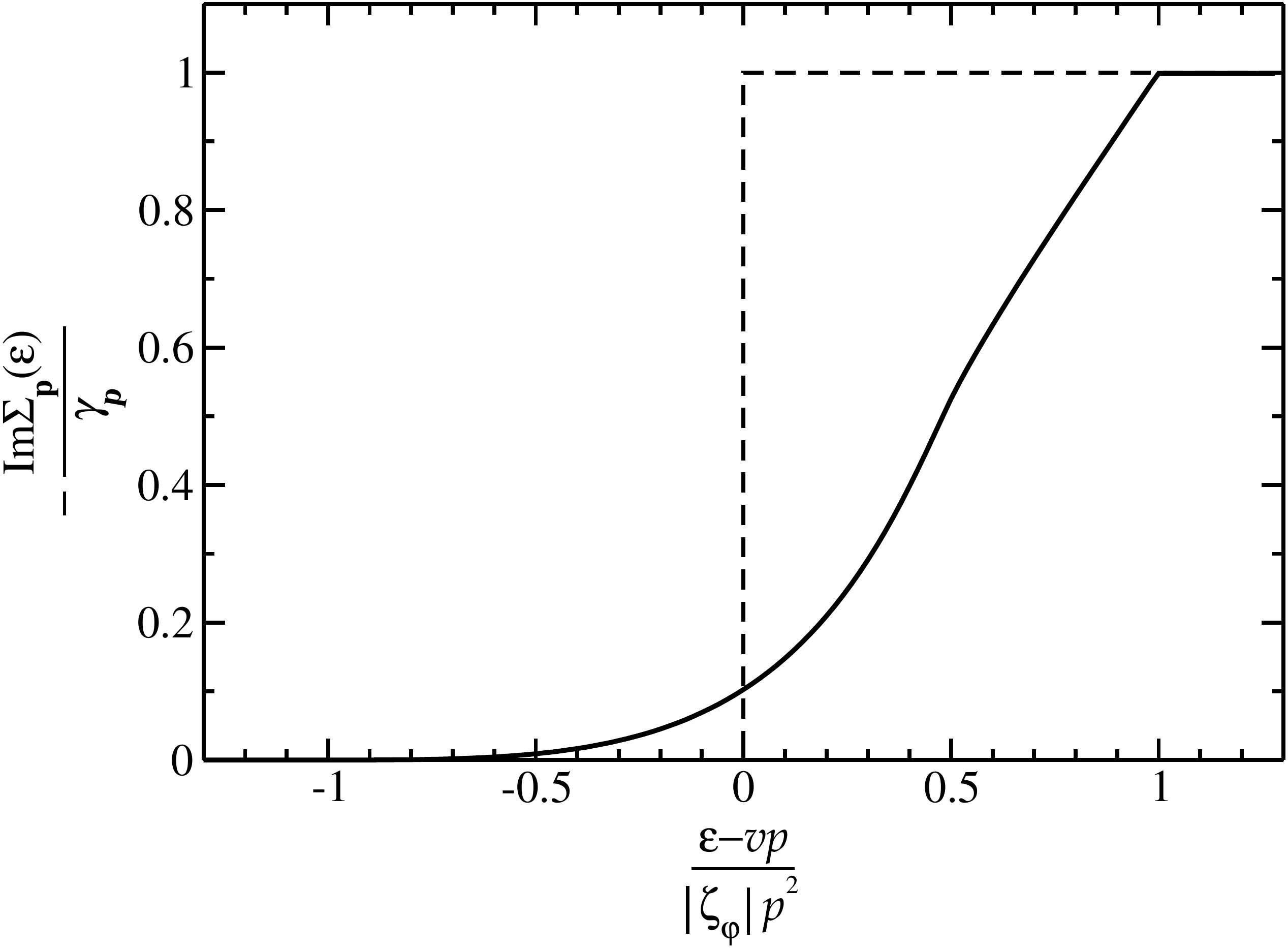}
\vspace{-0cm}
\end{center}
\caption{\label{fig:ImSigma2}
Smearing of the steplike discontinuity in the second-order
self-energy for $\zeta_\varphi>0$ at $N=4$. The dashed and
solid curves represent $-\Im\Sigma_\vec{p}(\ep)$ in and beyond
the Dirac approximation, respectively.
For $\zeta_\varphi<0$ the curve should be rotated by
$180^\circ$.}
\end{figure}

Beyond the Dirac approximation, we take into account the terms
proportional to $p^2$ in Eq.~(\ref{h=}). Assuming them to be
small compared to the main Dirac term~$vp$, we neglect them
wherever they produce small corrections to regular expressions
(e.~g., corrections to the eigenstates of the single-particle
Hamiltonian), and take them into account only in those terms
which are singular at $\ep\to{vp}$. In the second order of the
perturbation theory, instead of Eq.~(\ref{ImSigma2=}), we obtain
(see Appendix~\ref{app:ImSigma2trig} for details of the calculation)
\begin{equation}\begin{split}\label{ImSigma2trig=}
\zeta_\varphi>0:&\quad\Im\Sigma_\vec{p}(\ep)=-\gamma_\vec{p}\,\mathcal{F}_N\!
\left(\frac{\ep-vp}{\zeta_\varphi{p}^2}\right),\\
\zeta_\varphi<0:&\quad\Im\Sigma_\vec{p}(\ep)=-\gamma_\vec{p}\left[1-
\mathcal{F}_N\!\left(-\frac{\ep-vp}{|\zeta_\varphi|p^2}\right)\right],
\end{split}\end{equation}
where the function
$\mathcal{F}_N(z)$ is defined in
Appendix~\ref{app:ImSigma2trig}. The result of its numerical
evaluation for $N=4$ is plotted in Fig.~\ref{fig:ImSigma2}. 
$\mathcal{F}_N(z)=0$ for $z<-1$ .
On the mass shell, $\ep=vp+\zeta_\varphi{p}^2$, $\mathcal{F}_N=1$.
So, electronic relaxation is allowed if
$\cos{3}\varphi<-\zeta_3/\zeta_0\approx{0}.16$, with the rate
$2\gamma_\vec{p}$, and forbidden for
$\cos{3}\varphi>-\zeta_3/\zeta_0$. For a hole in the valence
band with momentum~$\vec{p}$, the conditions are just the
opposite.

Just like Eq.~(\ref{ImSigma2=}), Eq.~(\ref{ImSigma2trig=}) is
valid for $(Ne^2/v)^2vp\ll\ep-vp\ll{vp}$. When
$|\zeta_\varphi|p^2\lesssim(Ne^2/v)^2vp$ and $N\gg{1}$, the
self-energy should be calculated from RPA in the presence of
the $p^2$ terms. For this, one first has to calculate the
polarization operator $\Pi_\vec{q}(\omega)$ (the effect of
non-Dirac dispersion on the polarization operator in graphene
was also studied in Refs.~\onlinecite{Stauber2010,Stauber2011}).
Here we take the $p^2$ terms into account only near the
singularity at $\omega\to{v}q$, which becomes smeared as (see
Appendix~\ref{app:PolOpTrig} for details of the calculation)
\begin{subequations}\begin{eqnarray}
\label{Piq=}
&&\Pi_\vec{q}(\omega\approx{vq})=-\frac{Nq^2}{16\sqrt{2vq}}\,
\frac{1}{\sqrt{|\zeta_\varphi|q^2}}\,\mathcal{P}\!
\left(\frac{\omega-vq}{|\zeta_\varphi|q^2}\right),\\
&&\mathcal{P}(z)=\frac{8\sqrt{1-z}}{3\pi}
\left[(1+z)\,K\!\left(\frac{2}{1-z}\right)
-z\,E\!\left(\frac{2}{1-z}\right)\right],\nonumber\\
\label{Pz=}
\end{eqnarray}\end{subequations}
where $E(m)$ and $K(m)$ are the complete elliptic integrals.
The function $\mathcal{P}(z)$ is plotted in Fig.~\ref{fig:Pi}.
Finally, $\Im\Sigma_\vec{p}(\ep)$ is evaluated in
Appendix~\ref{app:ImSigmaRPAtrig}. It vanishes for
$\ep\leq{v}p-|\zeta_\varphi|p^2$. We give its explicit value
on the mass shell only, which has a compact form:
\begin{equation}
\label{ImSigmaRPAtrig=}
\Im\Sigma_\vec{p}(\ep^e_\vec{p})=
-\,\frac{64}{15\pi}\,\frac{\zeta_\varphi{p}^2}{N}\,\theta(\zeta_\varphi).
\end{equation}

\begin{figure}
\includegraphics[width=8cm]{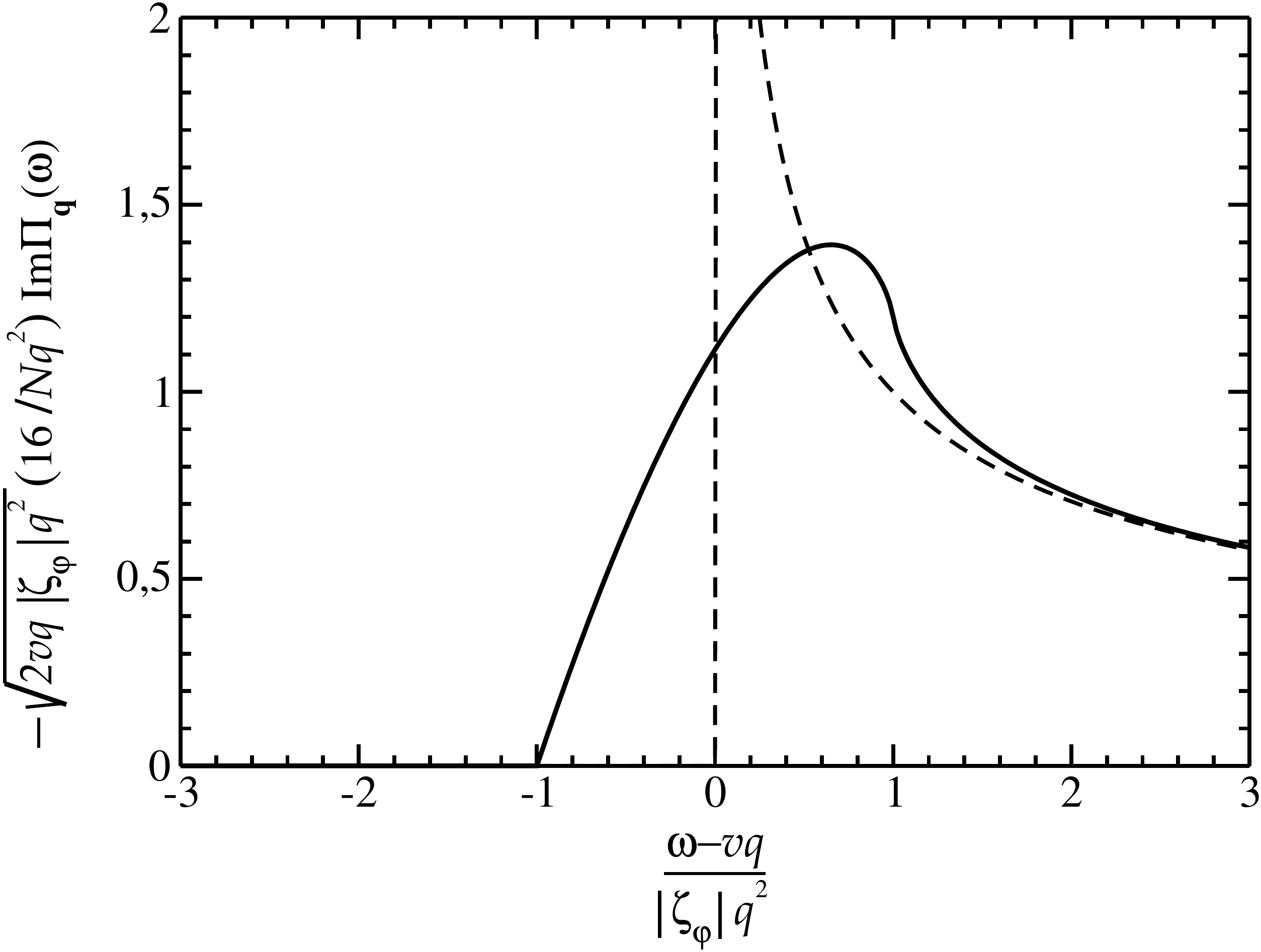}
\caption{\label{fig:Pi}
Smearing of the square root singularity at $\omega\to{vq}$
in the imaginary part of the polarization operator,
$-\Im\Pi_\vec{q}(\omega)$. The dashed curve represents the
Dirac approximation which gives $1/\sqrt{z}$, the solid curve
represents the function $\Im\mathcal{P}(z)$, defined in
Eq.~(\ref{Pz=}).}
\end{figure}

The above results [Eqs.~(\ref{ImSigma2trig=}) and
(\ref{ImSigmaRPAtrig=})] correspond to the Fermi golden gule
(FGR) with lowest-order or RPA-dressed transition matrix
elements. They are valid in the case
$|\zeta_\varphi|p^2\gg\min\{\gamma_\vec{p},vp/N\}$, when the
singularity in $\Im\Sigma_\vec{p}(\ep)$ is strongly smeared
by the band curvature. Let us study the opposite case,
when the dominant energy scale is $\gamma_\vec{p}$ itself,
e.~g., near the directions where $\zeta_\varphi=0$. It should
be recalled that FGR works only when the density of the final
states of the decay (three-particle excitations in the present
case) is approximately constant, in which case the quasiparticle
spectral peak has the Lorentzian shape. When the density of
states is not smooth, FGR-based approaches\cite{Rana2007,%
Winzer2010,Kim2011,Winzer2012,Winzer2013}
are not valid, and the quasiparticle spectral peak is manifestly
non-Lorentzian. To determine the quasiparticle properties in the
non-FGR regime, let us study the single-particle (retarded)
Green's function $G_\vec{p}(\ep)$, which determines the
quasiparticle spectral function, $-(1/\pi)\Im{G}_\vec{p}(\ep)$.

\begin{figure}
\begin{center}
\vspace{0cm}
\includegraphics[width=8cm]{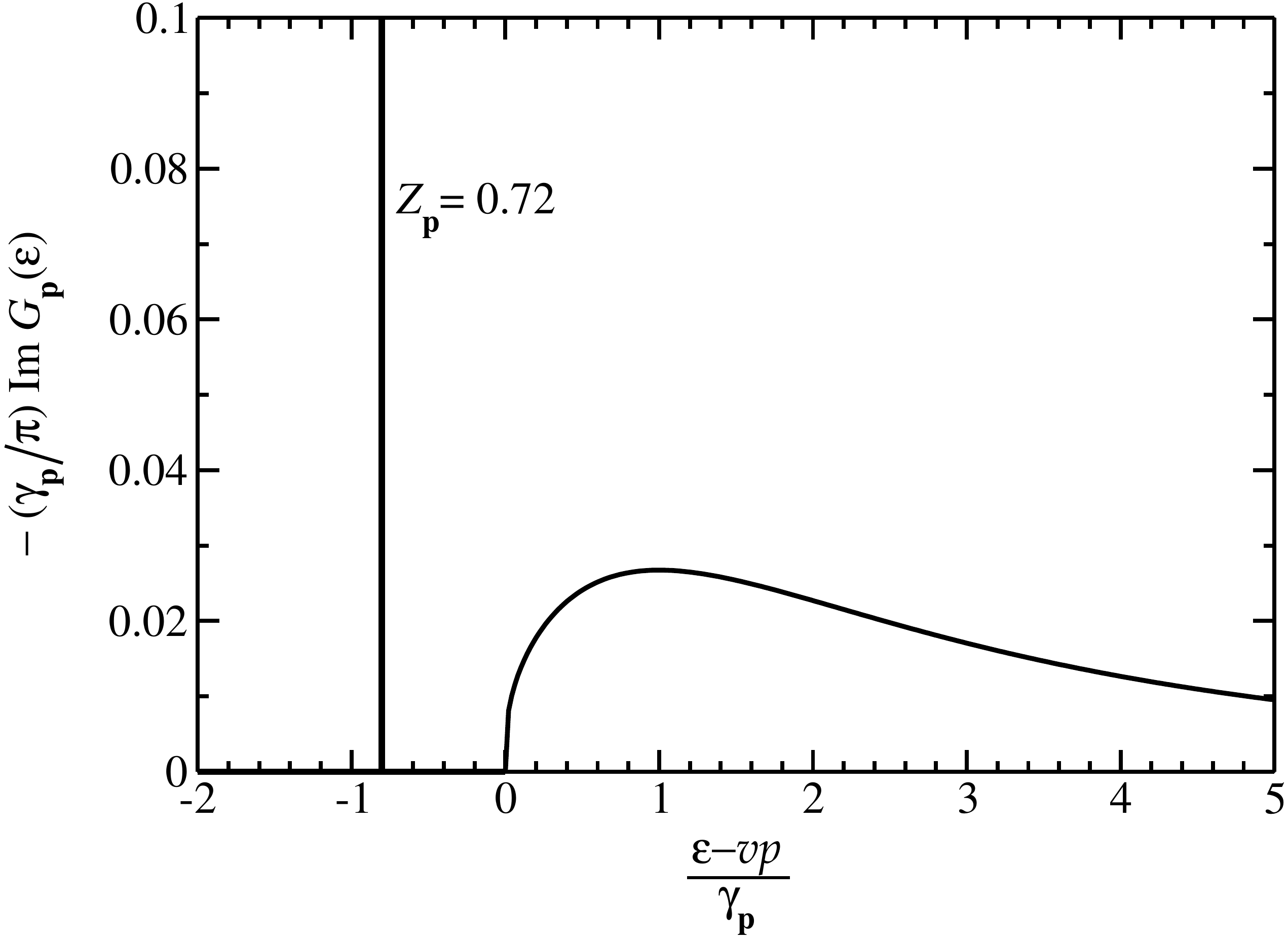}
\vspace{-0cm}
\end{center}
\caption{\label{fig:ImG}
The quasiparticle spectral function (in the units
of~$\gamma_\vec{p}$), as determined from Eq.~(\ref{Gp=})
for $\xi_\mathrm{max}/\gamma_\vec{p}=10$. The vertical peak
represents the $\delta$-function with the spectral weight
$Z_\vec{p}\approx{0}.72$.}
\end{figure}

Let us first analyze the most ``dramatic'' case when
$\Im\Sigma_\vec{p}(\ep)$ is given by Eq.~(\ref{ImSigma2=}).
In this case, the retarded Green's function (or, more precisely,
its projection on the eigenstate of the single-particle
Hamiltonian with momentum~$\vec{p}$) is given by
\begin{equation}\label{Gp=}
G_\vec{p}(\ep)=\left[\ep-vp
+\frac{\gamma_\vec{p}}\pi\ln\frac{\xi_\mathrm{max}}{|\ep-vp|}
+i\gamma_\vec{p}\theta(\ep-vp)\right]^{-1},
\end{equation}
with the real part of the self-energy reconstructed from the
Kramers-Kronig relation, and $\xi_\mathrm{max}\sim{vp}$
determines the ultraviolet cutoff of the logarithmic divergence,
since Eq.~(\ref{ImSigma2=}) is valid for $\ep-vp\ll{vp}$
only\cite{cutoff}. $G_\vec{p}(\ep)$ has a real pole at
$\ep-vp=-\Delta_\vec{p}<0$.
Its existence immediately follows from the fact that
$\Re\Sigma_\vec{p}(\ep)<0$, which, in turn, is a consequence of
the usual quantum-mechanical level repulsion: the quasiparticle
level is repelled from the three-particle continuum. 
Note that introduction of any infinitesimal broadening of the
step function does not affect this result at all.
The real pole corresponds to a quasiparticle state with an
infinite lifetime. Even though the quasiparticle does not decay
into the continuum, the latter still takes away part of the
spectral weight, which manifests itself in the residue
$Z_\vec{p}<1$ at the pole. With logarithmic precision we can
evaluate
\begin{equation}
\Delta_\vec{p}=\frac{\gamma_\vec{p}}\pi
\ln\frac{\xi_\mathrm{max}}{\gamma_\vec{p}},\quad
Z_\vec{p}=\frac{1}{1+1/\ln(\xi_\mathrm{max}/\gamma_\vec{p})}.
\end{equation}
(We remind that Eq.~(\ref{Gp=}) is valid only when
$\gamma_\vec{p}/\xi_\mathrm{max}\sim{N}(e^2/v)^2\ll{1}$, so
that the logarithm is large).
The spectral function for $\xi_\mathrm{max}/\gamma_\vec{p}=10$
is plotted in Fig.~\ref{fig:ImG}.

If now one gradually increases the band curvature
$\zeta_\varphi{p}^2>0$, (i)~the bare quasiparticle level is shifted
to $\ep=vp+\zeta_\varphi{p}^2$, (ii)~the spectral boundary of the
continuum is shifted to $\ep=vp-\zeta_\varphi{p}^2$, and (iii)~the
logarithmic divergence in $\Re\Sigma_\vec{p}(\ep\to{vp})$ at
$\ep\to{vp}$ is cut off at $\ep-vp\sim\zeta_\varphi{p}^2$.
The dressed quasiparticle level enters the continuum at some critical
value of $\zeta_\varphi{p}^2$, needed to overcome the level repulsion.
This critical value can be determined from the condition
$G_\vec{p}^{-1}(vp+\zeta_\varphi{p}^2)=0$, and with logarithmic
precision, it is given by
$\zeta_\varphi{p}^2=[\gamma_\vec{p}/(2\pi)]
\ln(\xi_\mathrm{max}/\gamma_\vec{p})$.
At this point the quasiparticle state acquires a finite
decay rate, whose value for sufficiently large
$\zeta_\varphi{p}^2$ is given by
$-2\Im\Sigma_\vec{p}(\ep_\vec{p}^e)$, as determined by
Eq.~(\ref{ImSigma2trig=}).

In RPA, suppression of $\Im\Sigma_\vec{p}(\ep)$,
Eq.~(\ref{ImSigmaRPA=}), at $\ep-vp\lesssim\xi_\mathrm{RPA}$ cuts
off the logarithmic divergence in $\Re\Sigma_\vec{p}(\ep\to{vp})$
at $\ep-vp\sim\xi_\mathrm{RPA}$ [here
$\xi_\mathrm{RPA}=\min\{vp,(Ne^2/v)^2vp\}$ is the same as in
Eq.~(\ref{ImSigmaRPA=})]. Since $\xi_\mathrm{RPA}\gg\gamma_\vec{p}$,
we can take $\Re\Sigma_\vec{p}(\ep=vp)$ to find the quasiparticle
pole,
\begin{subequations}
$\Delta_\vec{p}\approx-\Re\Sigma_\vec{p}(\ep=vp)$:
\begin{eqnarray}\label{DeltapRPAlog=}
&&\Delta_\vec{p}\approx\frac{\gamma_\vec{p}}\pi
\ln\frac{\xi_\mathrm{max}}{\xi_\mathrm{RPA}}\quad
(Ne^2/v\ll{1}),\\ \label{DeltapRPAsim=}
&&\Delta_\vec{p}\sim\frac{vp}N\quad (Ne^2/v\gg{1}),
\end{eqnarray}\end{subequations}
where $\xi_\mathrm{max}/\xi_\mathrm{RPA}\sim(Ne^2/v)^{-2}$.
When $Ne^2/v\ll{1}$, the logarithm is large, so
Eq.~(\ref{DeltapRPAlog=}) has logarithmic precision.
When $Ne^2/v\gg{1}$, Eq.~(\ref{DeltapRPAsim=}) represents
just an order-of-magnitude estimate obtained by plugging
Eq.~(\ref{ImSigmaRPA=}) into the Kramers-Kronig relation,
and integrating from $\ep-vp=0$ to $\ep-vp\sim{vp}$.
To find the residue at the pole, we note that at
$0<vp-\ep\ll\xi_\mathrm{RPA}$, from Eq.~(\ref{ImSigmaRPA=})
and the Kramers-Kronig relation one obtains
\begin{equation}
-\frac{\partial\Sigma_\vec{p}(\ep)}{\partial\ep}
\approx\frac{4}{\pi^2N}\int\limits_0^{\xi_\mathrm{RPA}}
\frac{\xi\ln(\xi_\mathrm{RPA}/\xi)\,d\xi}{(\xi+vp-\ep)^2}
\approx\frac{2}{\pi^2N}\ln^2\frac{\xi_\mathrm{RPA}}{vp-\ep},
\end{equation}
which gives
\begin{equation}
Z_\vec{p}=\left(1+\frac{2\ln^2N}{\pi^2N}\right)^{-1},
\end{equation}
with logarithmic precision.
The critical value of the band curvature, when the quasiparticle
level enters the continuum and acquires a finite decay rate is 
given by $\zeta_\varphi{p}^2\approx\Delta_\vec{p}/2$.

It is important that the existence of the infinitely narrow
quasiparticle peak in the Dirac approximation, obtained above
using the perturbation theory in $e^2/v$ and $1/N$ expansion,
is, in fact, more general that these approximations. Indeed,
the existence of the peak follows from two facts:
(i)~$\Im\Sigma_\vec{p}(\ep<vp)=0$, which holds in any order
of the perturbation theory because of energy and momentum
conservation as discussed in Appendix~\ref{app:ImSigma2}, and
(ii)~$\Re\Sigma_\vec{p}(\ep\to{vp})<0$, which is a consequence
of the level repulsion between the single-particle and the
three-particle states.
When the band curvature is included, the quasiparticle peak and
the continuum are pushed towards each other for $\zeta_\varphi>0$,
and away from each other for $\zeta_\varphi<0$.
Consequently, the requirement for the band curvature to exceed
a finite critical value in order to overcome the level repulsion
and to produce quasiparticle decay, is also more general than
the approximations used here. The calculation of critical value
itself, of course, does rely on approximations.

Still, one cannot exclude the appearance of nonzero
$\Im\Sigma_\vec{p}(\ep<vp)$ in the Dirac approximation due to
nonperturbative effects.
For example, nonperturbative generation of spectral weight
in $\Im\Pi_\vec{q}(\omega)$ at $\omega<vq$ due to excitonic
effects has been discussed in Ref.~\onlinecite{Mishchenko2008},
even though the validity of these results has been questioned in
Ref.~\onlinecite{Sodemann2012}. This issue calls for further
investigation.

\section{Discussion}

\begin{figure}
\begin{center}
\vspace{0cm}
\includegraphics[width=7cm]{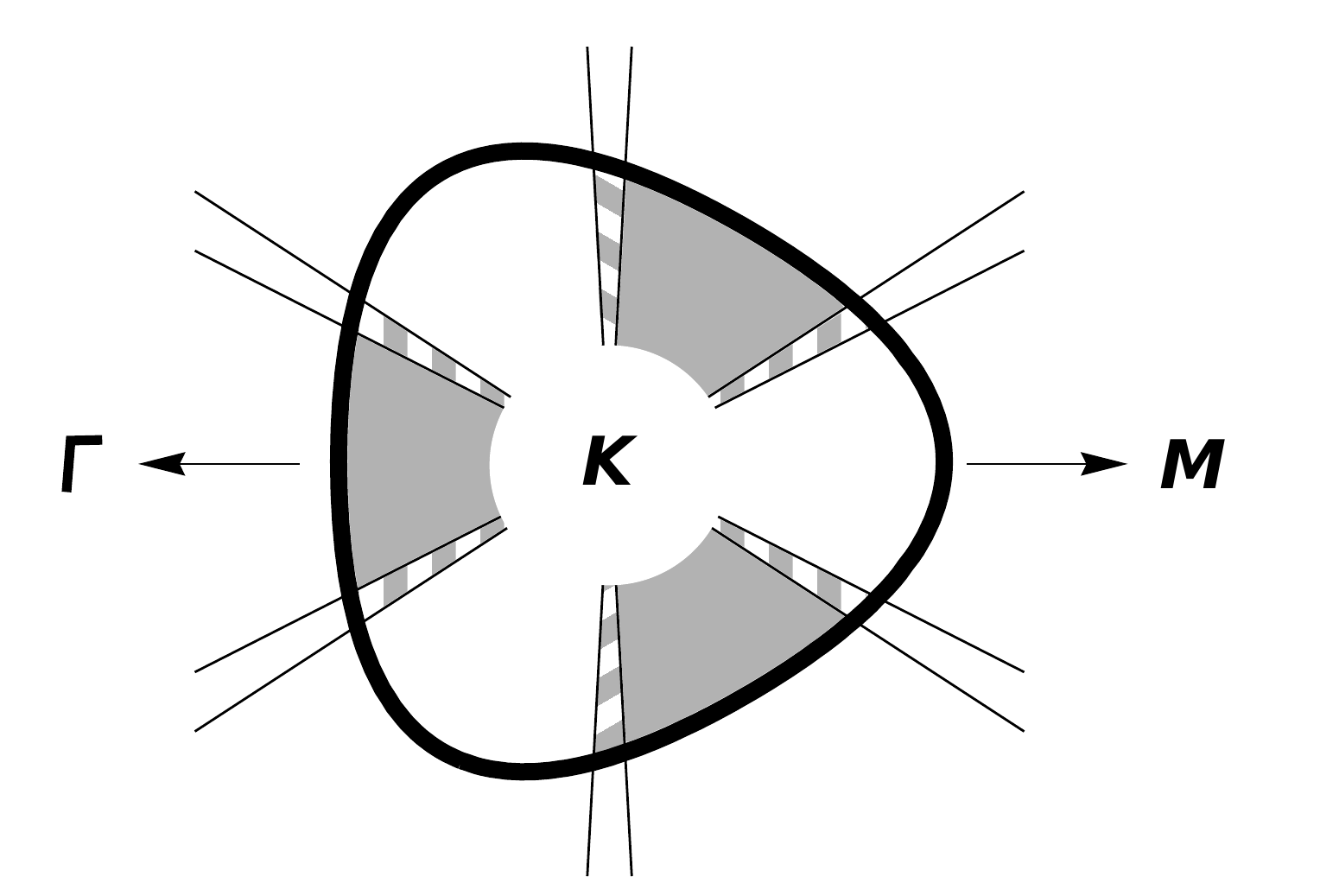}
\vspace{-0cm}
\end{center}
\caption{\label{fig:contours}
Directions of the electron momentum~$\vec{p}$ counted from the
$K$~point for which a
photoexcited e-h pair is subject to relaxation
by production of interband e-h pairs (carrier
multiplication). The thick solid curve represents the contour
$\ep^e_\vec{p}+\ep^h_{-\vec{p}}=2\:\mbox{eV}$, as determined
by Eq.~(\ref{epeh=}). The gray area shows the directions
of~$\vec{p}$ where both the electron with momentum~$\vec{p}$
and the hole with momentum $-\vec{p}$ are subject to
relaxation; along the directions in the hatched area, only
the hole is subject to relaxation.}
\end{figure}

Let us discuss some experimental implications of the obtained
results. In an optical experiment, the incident photon of the
frequency $\omega$ produces an electron with momentum $\vec{p}$
and a hole with momentum~$-\vec{p}$. Their energies satisfy
$\ep^e_\vec{p}+\ep^h_{-\vec{p}}=\omega$, which constrains
$\vec{p}$~to a trigonally warped circle, shown in
Fig.~\ref{fig:contours} for $\omega=2\:\mbox{eV}$.
The direction of~$\vec{p}$ is determined by the photon
polarization. If the excitation
density is low, one can neglect intraband collisions
between the photoexcited electrons and holes. According to
the above results, when the band curvature is sufficiently
large, the electron can relax by producing interband
e-h pairs (carrier multiplication) if the direction
of its momentum (the polar angle~$\varphi$) satisfies
$\zeta_\varphi=-\zeta_3\cos{3}\varphi-\zeta_0>0$ (shown in
Fig.~\ref{fig:contours} by the gray area). The hole can relax
if $-\zeta_{\pi-\varphi}=-\zeta_3\cos{3}\varphi+\zeta_0>0$
(gray and hatched areas on Fig.~\ref{fig:contours}).

The discussed anisotropy of the electronic relaxation has
some implications for the two-phonon Raman scattering, whose
intensity is suppressed by electronic
relaxation.\cite{Basko2008,Venezuela2011}. 
Namely, it favors the electronic states near the $KM$
direction (white sectors in Fig.~\ref{fig:contours}, the states
not subject to relaxation) to provide the dominant contribution
to the two-phonon Raman intensity, as has been observed
experimentally.\cite{Huang2010,Yoon2011}
Another mechanism favoring the electronic states near
the $KM$ direction is the anisotropy of the electron-phonon
coupling.\cite{Venezuela2011,Narula2012}

If the band curvature is too weak so that the quasiparticle
state does not decay, its spectral weight $Z_\vec{p}$ is still
reduced, $Z_\vec{p}<1$, since part of the spectral weight is
transferred to three-particle excitations.
It means that an initial excitation, produced by a short optical
pulse, has a finite probability $1-Z_\vec{p}$ to produce
many-particle excitations, the typical time of the processes
being $\sim{1}/\gamma_\vec{p}$ (since $\gamma_\vec{p}$~is the
typical energy scale of the features in the spectral function).
Thus, on average, the total number of e-h pairs per
absorbed photon will exceed unity, so one can still speak about
carrier multiplication even in the regime of weak band curvature.

\section{Acknowledgements}

The author is grateful to R. Asgari, I. V. Gornyi and M. Polini
for stimulating discussions.

\appendix

\section{Self-energy and polarization operator in the Dirac
approximation and beyond}

\subsection{General remarks about the calculation}
\label{app:general}

\begin{figure}
\begin{center}
\vspace{0cm}
\includegraphics[width=8cm]{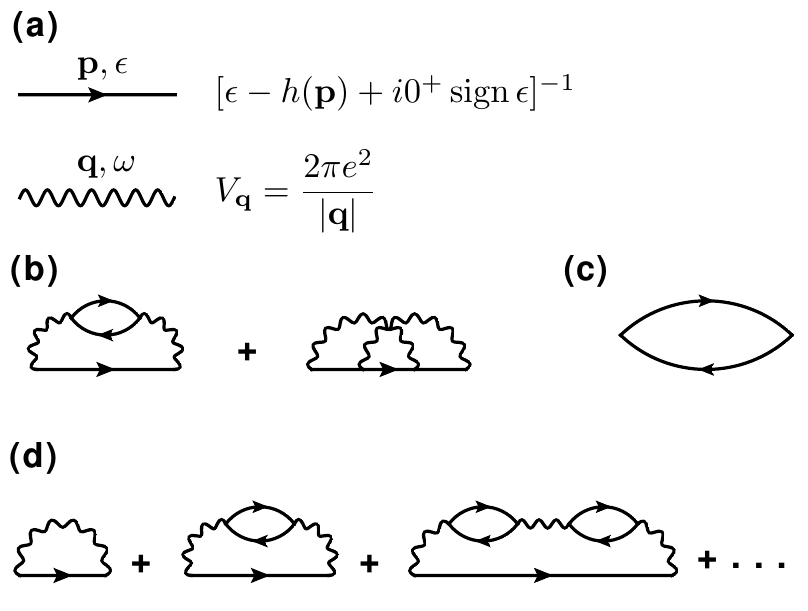}
\vspace{-0cm}
\end{center}
\caption{\label{fig:diagrams}
(a)~The basic elements of the diagrammatic technique 
(the solid line represents the electronic Green's function,
the wavy line represents the Coulomb interaction) and the
corresponding analytical expressions.
(b)~The two diagrams (direct and exchange) contributing to
$\Im\Sigma_\vec{p}(\ep)$ in the second order in $e^2/v$.
(c)~Electronic polarization operator.
(d)~The RPA series for the self-energy.
}
\end{figure}

The calculation is performed using the standard zero-temperature
diagrammatic technique whose basic elements, the single-particle
matrix
Green's function $[\ep-h(\vec{p})+i0^+\sign\ep]^{-1}$ with the
$2\times{2}$ matrix $h(\vec{p})$ given by Eq.~(\ref{h=}), and
the Coulomb interaction $V_\vec{q}=2\pi{e}^2/|\vec{q}|$, are
shown in Fig.~\ref{fig:diagrams}(a).

The self-energy is also a $2\times{2}$ matrix. In the Dirac
approximation, it can have components proportional to the unit
matrix or to the scalar product $\vec{p}\cdot\boldsymbol{\sigma}$,
due to isotropy of the problem.
Equivalently, the self-energy can be represented as
\[
\frac{|\vec{p}|+\vec{p}\cdot\boldsymbol{\sigma}}{2|\vec{p}|}\,
\Sigma_\vec{p}(\ep)
+\frac{|\vec{p}|-\vec{p}\cdot\boldsymbol{\sigma}}{2|\vec{p}|}\,
\bar\Sigma_\vec{p}(\ep),
\]
where $\Sigma_\vec{p}(\ep),\bar\Sigma_\vec{p}(\ep)$ are scalar
functions. The matrix coefficients in front of them are the
projectors on the two eigenstates of the Dirac Hamiltonian
with momentum~$\vec{p}$. Beyond the Dirac approximation, the
matrix structure of the self-energy becomes modified. However,
this modification represents a regular correction, proportional
to $\zeta_0,\zeta_3$, so it is neglected in all calculations,
since our primary interest is the singularity at $\ep\to{v}p$.

The singularities in the self-energy at $\ep\to{v}p$ and in
the polarization operator at $\omega\to{v}q$ come from nearly
collinear processes, i.~e., when all momenta are directed
almost along the same line. Thus, all angular factors resulting
from overlaps of eigenstates with different momenta, can be
dropped, as they produce small regular corrections to the main
singular behavior. This significantly simplifies the calculations.

\subsection{Second-order self-energy in the Dirac approximation}
\label{app:ImSigma2}

\begin{figure}
\begin{center}
\vspace{0cm}
\includegraphics[width=7cm]{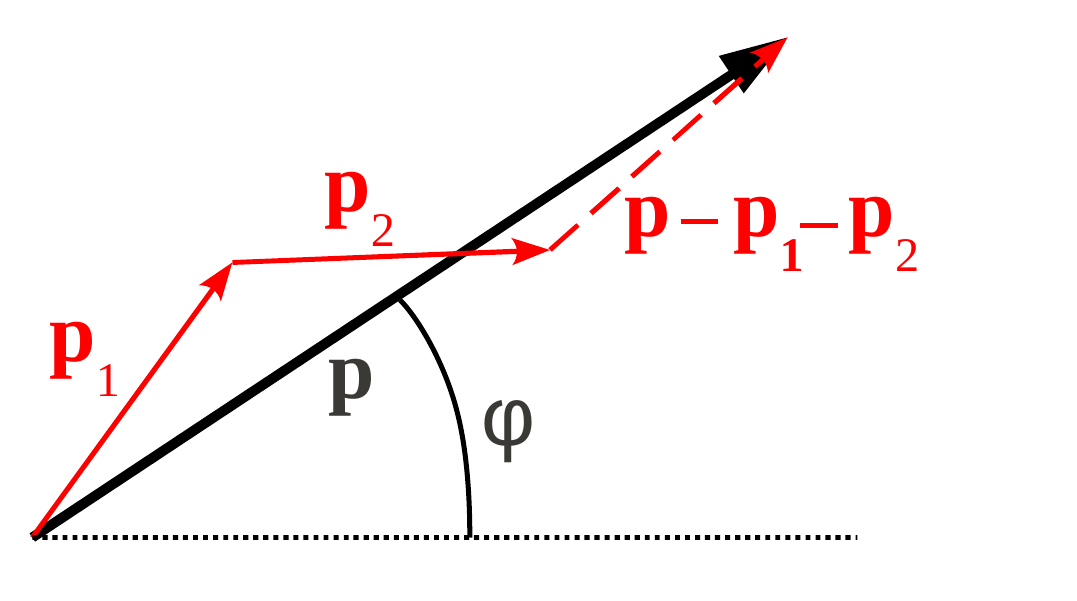}
\vspace{-0cm}
\end{center}
\caption{\label{fig:collinear}
The initial electron momentum~$\vec{p}$ (thick solid arrow) with
its polar angle~$\varphi$ with respect to the $KM$ direction
(thin dotted line),
and the final momenta of the two electrons $\vec{p}_1,\vec{p}_2$
(thin solid arrows) and of the hole $\vec{p}-\vec{p}_1-\vec{p}_2$
(thin dashed arrow).
}
\end{figure}

Upon integration over the internal energy variables, the sum of
the two diagrams in Fig.~\ref{fig:diagrams}(b) at $\ep\to{v}p$
can be written as
\begin{equation}\begin{split}
\Im\Sigma_\vec{p}(\ep)={}&{}-
\int\frac{d^2\vec{p}_1}{(2\pi)^2}\,\frac{d^2\vec{p}_2}{(2\pi)^2}\,
V_{\vec{p}-\vec{p}_1}
(NV_{\vec{p}-\vec{p}_1}-V_{\vec{p}-\vec{p}_2})\times\\
{}&{}\times\pi
\delta\!\left(\ep-\ep^e_{\vec{p}_1}-\ep^e_{\vec{p}_2}
-\ep^h_{\vec{p}-\vec{p}_1-\vec{p}_2}\right).
\label{ImS2epeeph=}
\end{split}\end{equation}
In the Dirac approximation $\ep^{e,h}_\vec{p}=v|\vec{p}|$, so
the $\delta$-function constrains $\ep/v$ to be equal to the
sum of the lengths of the three thin arrows in
Fig.~\ref{fig:collinear}. The triangle inequality ensures that
this is possible only when $\ep/v\geq{p}$, the length of the
long thick arrow. At $\ep\to{v}p$, the directions of
$\vec{p}_1,\vec{p}_2,\vec{p}-\vec{p}_1-\vec{p}_2$ should
approach the direction of~$\vec{p}$.

Let $x_1p,x_2p$ be the projections of $\vec{p}_1,\vec{p}_2$
on~$\vec{p}$, and $y_1p,y_2p$ the projections on the orthogonal
direction. 
The main contribution to the integral in Eq.~(\ref{ImS2epeeph=})
comes from the region $x_1,x_2>0$, $x_1+x_2<1$ and
$|y_1|,|y_2|\ll{1}$. Then, we can approximate
$|\vec{p}-\vec{p}_{1,2}|\approx{p}(1-x_{1,2})$ in the Coulomb
matrix elements, as they are nonsingular in the collinear limit
$y_{1,2}\to{0}$. In the energy $\delta$-function, $y_{1,2}$ are
kept to the second order:
\[\begin{split}
&\ep^e_{\vec{p}_{1,2}}\approx
{v}p\left(x_{1,2}+\frac{y_{1,2}^2}{2x_{1,2}}\right),\\
&\ep^h_{\vec{p}-\vec{p}_1-\vec{p}_2}\approx
{v}p\left[1-x_1-x_2+\frac{(y_1+y_2)^2}{2(1-x_1-x_2)}\right],
\end{split}\]
which gives
\[\begin{split}
&\Im\Sigma_\vec{p}(\ep)\approx
-\left(\frac{e^2}v\right)^2\frac{vp}{4\pi}{}\times{}\\
&\;\;{}\times{}\int\limits_0^1dx_1\int\limits_0^{1-x_1}dx_2
\left[\frac{N}{(1-x_1)^2}-\frac{1}{(1-x_1)(1-x_2)}\right]{}\times{}\\
&\;\;{}\times{}\int\limits_{-\infty}^\infty{d}y_1\,dy_2\,
\delta\!\left(\frac{\ep-vp}{vp}-\frac{y_1^2}{2x_1}-\frac{y_2^2}{2x_2}
-\frac{(y_1+y_2)^2/2}{1-x_1-x_2}\right).
\end{split}\]
The $y$-integration is performed using the general relation
\begin{equation}\begin{split}\label{detAintegral=}
&\int\limits_{-\infty}^\infty
\delta\!\left(s-\sum_{i,j=1}^{2n}A_{ij}y_iy_j\right)
\prod_{k=1}^{2n}dy_k=\\
&=\frac{1}{\sqrt{\det{A}}}\int\limits_{-\infty}^\infty
\delta\!\left(s-\sum_{i=1}^{2n}\tilde{y}_i^2\right)
\prod_{k=1}^{2n}d\tilde{y}_k=\\
&=\frac{\pi^ns^{n-1}\,\theta(s)}{\sqrt{\det{A}}\,(n-1)!},
\end{split}\end{equation}
valid for any positive-definite $2n\times{2n}$ matrix $A$.
The remaining $x$-integration,
\begin{equation}\begin{split}\label{ImS2x=}
&\Im\Sigma_\vec{p}(\ep)\approx
-\left(\frac{e^2}v\right)^2\frac{vp}2\,\theta(\ep-vp)
\int\limits_0^1dx_1\int\limits_0^{1-x_1}dx_2{}\times{}\\
&\times\sqrt{x_1x_2(1-x_1-x_2)}
\left[\frac{N}{(1-x_1)^2}-\frac{1}{(1-x_1)(1-x_2)}\right],
\end{split}\end{equation}
straightforwardly gives Eqs.~(\ref{ImSigma2=}),~(\ref{gamma=}).

Eq.~(\ref{detAintegral=}) also determines the suppression
by energy-momentum conservation of higher-order contributions
to $\Im\Sigma_\vec{p}(\ep\to{vp})$ corresponding to emission
of $n$~electron-hole pairs.
Indeed, this is precisely the kind of integral one obtains
for the perpendicular components of the $2n$ momenta in the
collinear limit, with $s\propto(\ep-vp)$.

\subsection{Second-order self-energy beyond the Dirac approximation}
\label{app:ImSigma2trig}

Neglecting the trigonal warping correction to the single-particle
eigenstates, one can again use Eq.~(\ref{ImS2epeeph=}) with the
quasiparticle dispersions from Eq.~(\ref{epeh=}) to calculate the
self-energy. Using the same notation as in
Appendix~\ref{app:ImSigma2}, we expand the quasiparticle dispersions
to the second
order in $y_1,y_2$. Consider, for example, $\ep^e_{\vec{p}_1}$:
\begin{equation}\begin{split}\label{epexy=}
\ep^e_{\vec{p}_1}={}&{}vpx_1-(\zeta_0+\zeta_3\cos{3}\varphi)p^2x_1^2
+3\zeta_3p^2x_1y_1\sin{3}\varphi{}+{}\\
&{}+{}\left(\frac{vp}{2x_1}-\zeta_0p^2
+\frac{7}{2}\,\zeta_3p^2\cos{3}\varphi\right)y_1^2.
\end{split}\end{equation}
The expression for $\ep^h_{\vec{p}-\vec{p}_1-\vec{p}_2}$ has a
similar structure. The terms linear in $y_1,y_2$ can be removed
by a shift of $y_1,y_2$. Neglecting the terms
$\sim(\zeta_3p^2)^2/(vp)$ as well as corrections to the
determinant in Eq.~(\ref{detAintegral=}), we obtain the same
Eq.~(\ref{ImS2x=}), but with a modified integration domain:
in addition to the conditions $x_1,x_2>0$, $x_1+x_2<1$,
there is another condition
\begin{equation}\label{condition=}
\ep-vp-\zeta_\varphi{p}^2+2\zeta_\varphi{p}^2(1-x_1)(1-x_2)>0,
\end{equation}
where $\zeta_\varphi=-\zeta_0-\zeta_3\cos{3}\varphi$.
Even though we could not evaluate the corresponding integral 
analytically, some general properties of it can be established.
First, for sufficiently large negative $\ep-vp<-|\zeta_\varphi|p^2$
the integration domain is empty, so $\Im\Sigma_\vec{p}(\ep)=0$.
Second, for sufficiently large positive $\ep-vp>|\zeta_\varphi|p^2$,
the condition~(\ref{condition=}) becomes redundant, the integration
domain coincides with that in Eq.~(\ref{ImS2x=}), so
$\Im\Sigma_\vec{p}(\ep)=\gamma_\vec{p}$.
Finally, the condition~(\ref{condition=}) becomes its opposite upon
simultaneous change of sign of $\ep-vp$ and $\zeta_\varphi$. Thus,
if one defines a function $\mathcal{F}_N(z)$ as
\begin{subequations}\begin{eqnarray}
&&\mathcal{F}_N(z)=\left[\int\mathcal{I}_N(x_1,x_2)\,dx_1\,dx_2\right]^{-1}
\times\nonumber\\
&&\quad{}\times{}\int\theta(z-1+2(1-x_1)(1-x_2))\,
\mathcal{I}_N(x_1,x_2)\,dx_1\,dx_2,\nonumber\\
&&\\
&&\mathcal{I}_N(x_1,x_2)\equiv
\theta(x_1)\,\theta(x_2)\,\theta(1-x_1-x_2)\,
\times\nonumber\\
&&\qquad{}\times{}\sqrt{x_1x_2(1-x_1-x_2)}{}\times{}\nonumber\\
&&\qquad{}\times{}
\left[\frac{N}{(1-x_1)^2}-\frac{1}{(1-x_1)(1-x_2)}\right],
\end{eqnarray}\end{subequations}
then
$\Im\Sigma_\vec{p}(\ep)$ is given by Eq.~(\ref{ImSigma2trig=}).

\subsection{Polarization operator in the Dirac approximation}
\label{app:PolOp}

The exact expression for the polarization operator in the Dirac
approximation is known since long ago\cite{Gonzalez1994}:
\begin{equation}\label{Pistandard=}
\Pi_\vec{q}(\omega)=-\frac{Nq^2/16}{\sqrt{v^2q^2-\omega^2}}.
\end{equation}
Still, here we give its simple derivation near the singularity
at $\omega\to{vq}$, which will be generalized beyond the Dirac
approximation in Sec.~\ref{app:PolOpTrig}. In the collinear
approximation the angular factors can be neglected, so the
polarization operator is given by
\begin{equation}\label{Piepeeph=}
\Pi_\vec{q}(\omega)\approx\int\frac{d^2\vec{p}}{(2\pi)^2}\,
\frac{N}{\omega-\ep^e_\vec{p}-\ep^h_{\vec{q}-\vec{p}}}.
\end{equation}
Let us denote by $xq$ the projection of $\vec{p}$ on~$\vec{q}$.
and by $yq$ on the orthogonal direction. Then,
\[\begin{split}
&\Pi_\vec{q}(\omega)\approx-\frac{Nq}{4\pi^2v}
\int\limits_{-\infty}^\infty{d}x\,dy\times\\
&\qquad{}\times\left(|x|+|1-x|+\frac{y^2}{2|x|}+\frac{y^2}{2|1-x|}
-\frac\omega{vq}\right)^{-1}=\\
&=-\frac{Nq}{4\pi{v}}\int\limits_{-\infty}^\infty
\frac{\sqrt{2|x(1-x)|}\,dx}%
{\sqrt{(|x|+|1-x|)[|x|+|1-x|-\omega/(vq)]}}=\\
&=-\frac{Nq^2}{2\pi\sqrt{2vq(vq-\omega)}}
\int\limits_0^1\sqrt{x(1-x)}\,dx{}-{}\\
&\qquad{}-{}\frac{Nq}{2\pi{v}}\int\limits_{1/2}^\infty
\frac{\sqrt{u^2-1/4}\,du}{\sqrt{u[2u-\omega/(vq)]}}.
\end{split}\]
The second term is nonsingular at $\omega\to{v}q$, so it can
be ignored (the divergence of the integral is spurious, being
a consequence of the collinear approximation).
The first one gives $-(Nq^2/16)/\sqrt{2vq(vq-\omega)}$, which
is precisely Eq.~(\ref{Pistandard=}) at $\omega\to{vq}$.

\subsection{Polarization operator beyond the Dirac approximation}
\label{app:PolOpTrig}

As in Sec.~\ref{app:PolOp}, let us start from Eq.~(\ref{Piepeeph=})
and denote by $xq$ the projection of $\vec{p}$ on~$\vec{q}$, and
by $yq$ on the orthogonal direction.
As in Sec.~\ref{app:ImSigma2trig}, let us expand the energies
to the second order in~$y$, see Eq.~(\ref{epexy=}). In the interval
$0<x<1$, which determines the main singularity, we have
\[
\ep^e_\vec{p}+\ep^h_{\vec{q}-\vec{p}}\approx
vq+\zeta_\varphi{q}^2(2x-1)
+\frac{vq}{2}\left(\frac{1}{x}+\frac{1}{1-x}\right)(y-y_0)^2,
\]
where $y_0$ is an $x$-dependent shift. The integration over~$y$ is
straightforward, the subsequent integral over~$x$ reduces to
elliptic integrals, to give Eqs.~(\ref{Piq=}),~(\ref{Pz=}).

Let us mention some properties of the function~$\mathcal{P}(z)$,
defined in Eq.~(\ref{Pz=}). Many of them can be deduced directly
from the integral representation,
\begin{equation}\label{calPzint=}
\mathcal{P}(z)=
\frac{2}\pi\int\limits_{-1}^1\sqrt{\frac{1-u^2}{u-z}}\,du.
\end{equation}
For $z<-1$, $\mathcal{P}(z)$ is purely real, at $z>-1$ it
acquires a positive imaginary part whose sign is fixed by the
requirement of the analyticity in the upper half-plane of
the complex variable~$z$, and $\mathcal{P}(z)$ is purely
imaginary at $z>1$. The real and imaginary parts are related
by $\Re\mathcal{F}(-z)=\Im\mathcal{F}(z)$. At $z\to-\infty$,
$\mathcal{P}(z)\to{1}/\sqrt{-z}$. At $z\to\pm{1}$,
$\mathcal{P}(z)$ has a weak singularity:
\begin{equation}\label{calPz-1=}
\mathcal{P}(z)\approx\frac{8\sqrt{2}}{3\pi}
+\frac{\sqrt{2}\,(z+1)}{\pi}
\left[\ln\frac{A}{|z+1|}+i\pi\,\theta(z+1)\right],
\end{equation}
with $A\sim{1}$. Finally, there are two integral relations,
valid at $y>1$:
\begin{subequations}\begin{eqnarray}
&&\int\limits_{-1}^{y}\frac{dz}{\sqrt{y-z}}\,\Im\mathcal{P}(z)=\pi,\\
&&\int\limits_{-1}^{y}\frac{dz}{\sqrt{y-z}}\,
\frac{\Im\mathcal{P}(z)}{|\mathcal{P}(z)|^2}=\frac{\pi{y}}{2}.\label{int1calP=}
\end{eqnarray}\end{subequations}
The first relation can be proven by using Eq.~(\ref{calPzint=})
and interchanging the order of integration, while the second
relation has been verified numerically.

\subsection{RPA self-energy in the Dirac approximation}
\label{app:ImSigmaRPA}

Upon integration over the internal frequency variable of the
RPA diagrams, the self-energy becomes
\begin{equation}\label{ImSigmaVP=}
\Im\Sigma_\vec{p}(\ep)\approx\int\frac{d^2\vec{q}}{(2\pi)^2}\,
V_\vec{q}^2\Im\frac{\Pi_\vec{q}(\ep-\ep^e_{\vec{p}-\vec{q}})}%
{1-V_\vec{q}\Pi_\vec{q}(\ep-\ep^e_{\vec{p}-\vec{q}})}.
\end{equation}
The imaginary part of the dressed polarization operator is given by
\begin{subequations}\begin{eqnarray}
&&\Im\frac{\Pi_\vec{q}(\omega)}{1-V_\vec{q}\Pi_\vec{q}(\omega)}
=-\frac{Nq^2}{16}\,\frac{\sqrt{\omega^2-v^2q^2}}{\omega^2-\tilde{v}^2q^2}\,
\theta(\omega-vq),\nonumber\\ &&\\
&&\frac{\tilde{v}^2}{v^2}=1-\left(\frac{\pi}{8}\,\frac{Ne^2}{v}\right)^2.
\end{eqnarray}\end{subequations}
Note that $\tilde{v}^2$ can become negative, which does not
represent any problem; the main effect of $\tilde{v}^2$ is to
make the denominator nonzero at $\omega\to{v}q$. Because
of this, in the resulting expression for the self-energy,
\begin{equation}\begin{split}
\Im\Sigma_\vec{p}(\ep)\approx{}&{}
-\frac{Ne^4}{16}\int{d}^2\vec{q}\,\theta(\ep-v|\vec{p}-\vec{q}|-vq)\times{}\\
{}&{}\times\frac{\sqrt{(\ep-v|\vec{p}-\vec{q}|)^2-v^2q^2}}%
{(\ep-v|\vec{p}-\vec{q}|)^2-\tilde{v}^2q^2},
\end{split}\end{equation}
one can simply set $\ep=vp$ in the denominator and replace the
latter by $(v^2-\tilde{v}^2)q^2$ provided that
$|\ep-vp|\ll(Ne^2/v)^2vq$.
This condition will provide a lower cutoff for the integration
over~$q$.

In the numerator, we approximate
\[
(\ep-v|\vec{p}-\vec{q}|)^2-v^2q^2\approx
2vq_\|(\ep-vp)-\frac{p(vq_\perp)^2}{p-q_\|},
\]
where $q_\|,q_\perp$ are the components of $\vec{q}$
along~$\vec{p}$ and perpendicular to~$\vec{p}$, respectively,
and $|q_\perp|\ll{q}_\|$.
Then after the straightforward integration over $q_\perp$,
we obtain
\begin{equation}\begin{split}
\Im\Sigma_\vec{p}(\ep)={}&{}
\frac{4(\ep-vp)}{\pi{N}}\int\limits_0^p\frac{dq_\|}{q_\|}\,
\sqrt{\frac{p-q_\|}{p}}\approx\\
\approx{}&{}\frac{4(\ep-vp)}{\pi{N}}\,\ln\frac{p}{q_{min}},
\end{split}\end{equation}
where the logarithmic divergence is cut off at
\[
vq_{min}=\max\left\{\ep-vp,\frac{\ep-vp}{(Ne^2/v)^2}\right\},
\]
as discussed above. Thus, we arrive at Eq.~(\ref{ImSigmaRPA=}).

\subsection{RPA self-energy beyond the Dirac approximation}
\label{app:ImSigmaRPAtrig}

Let us again start from the general Eq.~(\ref{ImSigmaVP=}). Beyond
the Dirac approximation, we use Eq.~(\ref{Piq=}) and obtain
\begin{subequations}
\begin{equation}\label{ImPiVPi=}
\Im\frac{V_\vec{q}^2\Pi_\vec{q}(\ep-\ep^e_{\vec{p}-\vec{q}})}%
{1-V_\vec{q}\Pi_\vec{q}(\ep-\ep^e_{\vec{p}-\vec{q}})}
\approx\frac{2\pi{e}^2}{q_\|}\,\frac{\sqrt{\mathcal{A}}\Im\mathcal{P}(z)}%
{|1+\sqrt{\mathcal{A}}\,\mathcal{P}(z)|^2},
\end{equation}
where we denoted
\begin{eqnarray}
&&\mathcal{A}=\left(\frac{\pi{N}e^2}{8v}\right)^2
\frac{vq_\|}{2|\zeta_\varphi|q_\|^2},\\
&&z=\frac{1}{|\zeta_\varphi|q_\|^2}
\left[\ep-vp-\frac{vpq_\perp^2}{2q_\|(p-q_\|)}-\zeta_\varphi(p-q_\|)^2\right].
\nonumber\\ &&
\end{eqnarray}\end{subequations}
In the relevant region of energies, namely, where
Eq.~(\ref{ImSigmaRPA=}) is valid in the Dirac approximation,
we have $\mathcal{A}\gg{1}$ and $z\ll\mathcal{A}$. Due to
the latter condition and to the fact that
$\Im\mathcal{P}(z)\sim{1}/\sqrt{z}$ at $z\to\infty$, one
can neglect the unity in the denominator of Eq.~(\ref{ImPiVPi=}),
which then becomes
\begin{equation}
\Im\frac{V_\vec{q}^2\Pi_\vec{q}(\ep-\ep^e_{\vec{p}-\vec{q}})}%
{1-V_\vec{q}\Pi_\vec{q}(\ep-\ep^e_{\vec{p}-\vec{q}})}
\approx\frac{16}{N}\sqrt{\frac{2|\zeta_\varphi|v}{q_\|}}\,
\frac{\Im\mathcal{P}(z)}{|\mathcal{P}(z)|^2}.
\end{equation}
Let us pass from the integration over $q_\perp$ to the one
over~$z$, 
which gives
\begin{subequations}\begin{eqnarray}\label{ImSxz=}
&&\Im\Sigma_\vec{p}(\ep)=\frac{8|\zeta_\varphi|p^2}{\pi^2N}
\int\limits_0^1x\sqrt{1-x}\,dx\times\nonumber\\
&&\qquad{}\times\int\limits_{-1}^{z_{max}}
\frac{\theta(z_{max}+1)\,dz}{\sqrt{z_{max}-z}}\,
\frac{\Im\mathcal{P}(z)}{|\mathcal{P}(z)|^2},\\
&&z_{max}(x)=\frac{\ep-vp-\zeta_{\varphi}p^2(1-x)^2}{|\zeta_\varphi|p^2x^2}.
\end{eqnarray}\end{subequations}
The $z$-integration is performed using Eq.~(\ref{int1calP=}).
On the mass shell, $\ep=vp+\zeta_\varphi{p}^2$, the $x$-integral
in Eq.~(\ref{ImSxz=}) converges on the lower limit, and one arrives
at Eq.~(\ref{ImSigmaRPAtrig=}).


\end{document}